\pdfoutput=1

\documentclass[galaxies,article,accept,moreauthors,10pt,a4paper]{mdpi} 

\usepackage{listings}
\firstpage{1} 
\pubvolume{xx}
\issuenum{1}
\articlenumber{1}
\pubyear{2018}
\copyrightyear{2018}
\history{Received: date; Accepted: date; Published: date}

\Title{Science Pipelines for the Square Kilometre Array}

\Author{Jamie Farnes$^{1}$\orcidA{}*, Ben Mort$^{1}$, Fred Dulwich$^{1}$, Stef Salvini$^{1}$, Wes Armour$^{1}$}

\AuthorNames{Firstname Lastname, Firstname Lastname and Firstname Lastname}

\address{%
$^{1}$ \quad Oxford e-Research Centre (OeRC), Department of Engineering Science, University of Oxford, Oxford, OX1 3QG, UK.}

\corres{Corresponding author: jamie.farnes@oerc.ox.ac.uk}

\abstract{The Square Kilometre Array (SKA) will be both the largest radio telescope ever constructed and the largest Big Data project in the known Universe. The first phase of the project will generate on the order of 5~zettabytes of data per year. A critical task for the SKA will be its ability to process data for science, which will need to be conducted by science pipelines. Together with polarization data from the LOFAR Multifrequency Snapshot Sky Survey (MSSS), we have been developing a realistic SKA-like science pipeline that can handle the large data volumes generated by LOFAR at 150~MHz. The pipeline uses task-based parallelism to image, detect sources, and perform Faraday Tomography across the entire LOFAR sky. The project thereby provides a unique opportunity to contribute to the technological development of the SKA telescope, while simultaneously enabling cutting-edge scientific results. In this paper, we provide an update on current efforts to develop a science pipeline that can enable tight constraints on the magnetised large-scale structure of the Universe.}


\begin{document}

\section{Introduction}
The Square Kilometre Array (SKA) will be the largest radio telescope ever constructed. The completed telescope will span two continents -- with the project being jointly located in South Africa, where hundreds of dishes will be built, and in Australia, where 100,000 dipole antennas will be situated. Through the technique of interferometry, the signals from these antennas can be combined so that the array behaves like a single dish that is 150~km in diameter.

The SKA Organisation itself is located next to Jodrell Bank Observatory in Manchester, with the UK community being led by Manchester, Oxford, and Cambridge, together with UK industrial partners. The UK, via the Science and Technology Facilities Council (STFC), are contributing $\pounds$100M towards construction of the first phase of the project. The key participating nations within the project include Australia, Canada, China, France, India, Italy, New Zealand, South Africa, Spain, Sweden, the Netherlands, and the UK. Further countries have expressed their interest in joining the SKA Organisation, which will continue to expand over the coming years.

\begin{figure}[H]
\centering
\includegraphics[width=8.3 cm]{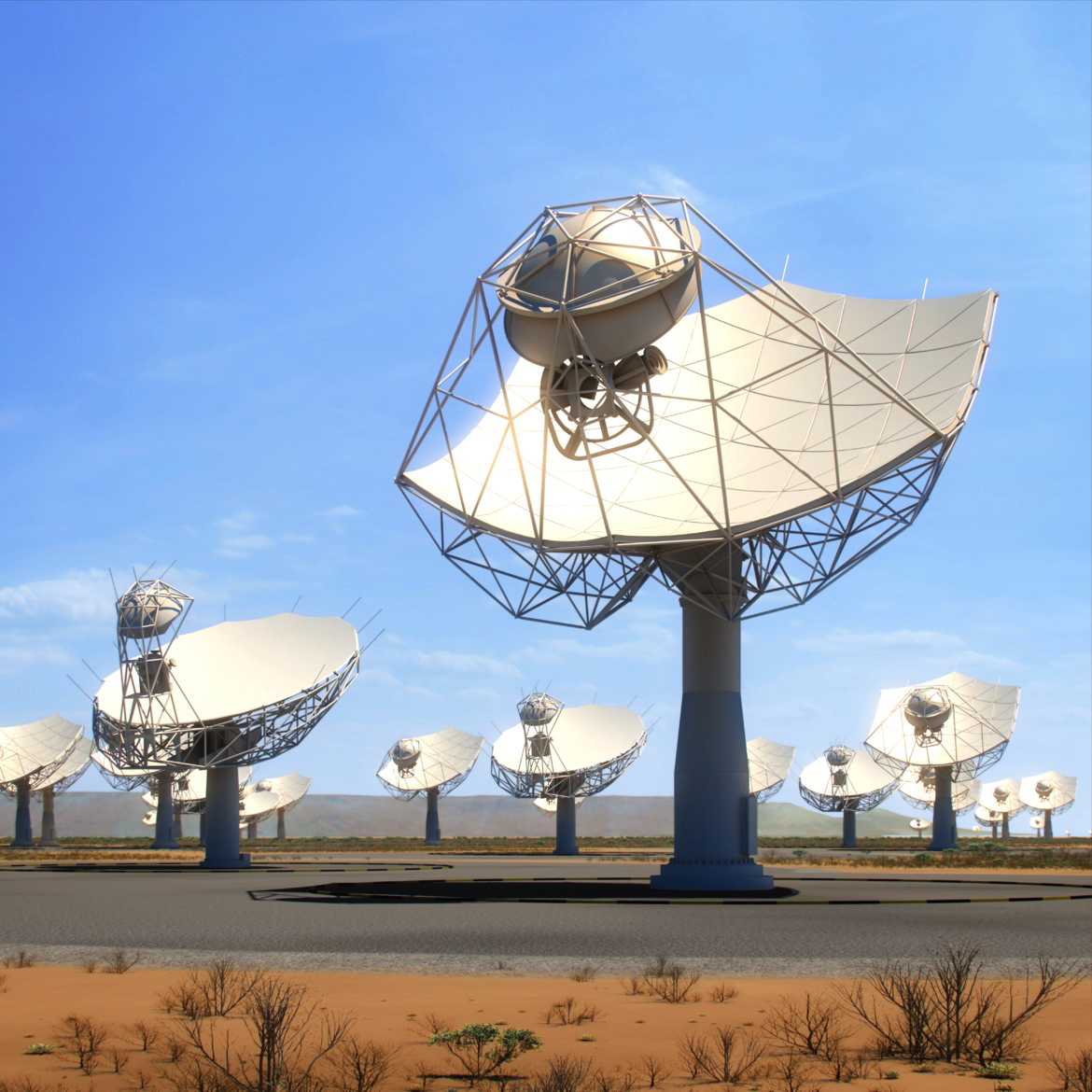}
\includegraphics[width=8.3 cm]{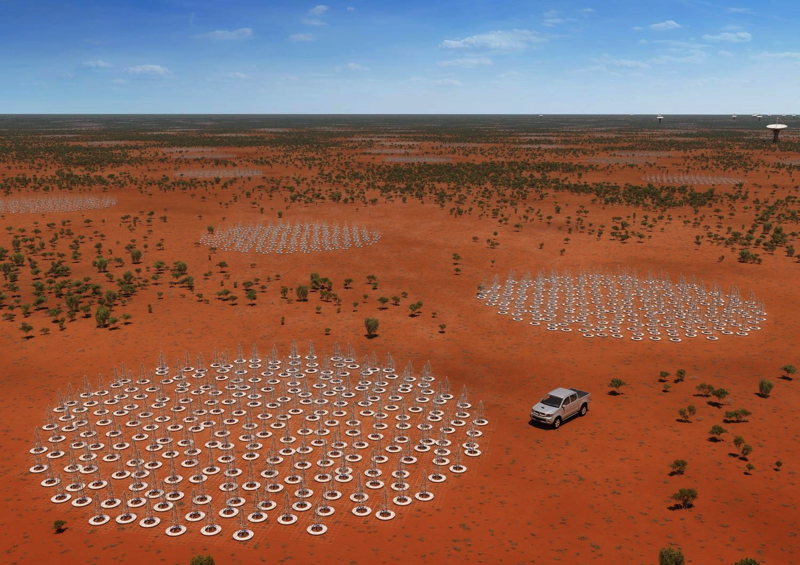}
\caption{Top: The dishes for SKA-Mid, located in South Africa, will generate 62~exabytes of raw data per year, which is enough to fill 340,000~average laptops with content every day. Bottom: The aperture arrays for SKA-Low, located in Australia, will generate 5~zettabytes of data per year. Unlike conventional dishes, the ``beam'' of these antennas is formed electronically -- allowing each station to look at multiple regions of the sky simultaneously. The total field of view is essentially limited by the signal processing capacity. Image Credits: SKA Organisation.}
\label{SKA}
\end{figure}

Both of the SKA telescope sites are ideal locations in which to situate a radio telescope, see Fig.~\ref{SKA}. The Australian site is located in the Western Australian outback, approximately 800~km from Perth, within a government-protected radio quiet zone. This radio quiet zone spans an area of $\approx70,685$~km$^2$, which is approximately equivalent to twice the area of the island of Kyushu in Japan, or about half the area of England in the UK, but with a population of less than $\sim100$~people. This remote location is highly advantageous for radio astronomy, as humans are radio-loud with mobile phones, microwaves, and various other emitters of radio frequency interference (RFI). All of these modern radio transmitters can dominate the tiny radio signals to which the telescope is sensitive and that have travelled from the other side of the Universe.

The term ``Big Data'' is often problematic, in that it is highly subjective. However, the SKA will undoubtedly be the largest Big Data project in the known Universe. The first phase of the project will generate on the order of exabytes ($=1,000$~petabytes) of raw data from the antennas every day. The Australian counterpart of the project alone -- SKA-Low -- will generate 157~terabytes of data every second. Overall, this amounts to $\approx$5~zettabytes ($10^6$~petabytes) of data that will be generated by the facility each year throughout the first phase of the project. This latter data rate is equivalent to $5\times$ the estimated global internet traffic in 2015.\footnote{As estimated by the Cisco Systems Visual Networking Index (VNI).}

This data flood from the SKA will have implications for medical researchers, neuroscientists, data scientists, quantitative financial analysts, and the general public throughout the global community, in terms of how we handle and process such large datasets. The infrastructure for handling these data rates does not currently exist. To some extent, new technologies will enable us to process the data when the facility comes online in 2023. However, new techniques, algorithms, and infrastructure also need to be developed in order to solve these data challenges. In this paper, we discuss current efforts at developing these new techniques and our contribution to various aspects of SKA development.

\section{Magnetism Surveys as Prototyping Test-Beds}

The science case for the SKA -- a state-of-the-art radio telescope -- covers a diverse range of topics which includes challenging questions such as: how do galaxies evolve?, what is the nature of Dark energy?, was Einstein right about gravity?, what generates giant magnetic fields in space?, and even whether we are alone in the Cosmos? \cite{b1}. For the purpose of prototyping, SKA-like science pipelines could be developed to cover the technical requirements for any of these areas. We have here chosen to focus on the science case of magnetic fields, which can be advantageous from a prototyping perspective as the data processing requires full-Stokes polarization imaging in $I$, $Q$, $U$, and $V$ \cite{1996A&AS..117..137H}, together with Faraday Tomography \cite{1966MNRAS.133...67B,2014PASJ...66...61K,2016arXiv160301974A,2018PASJ...70R...2A}, which intrinsically has substantial computing requirements. The Big Data challenges associated with Faraday Tomography will be further increased at SKA sensitivities, for which a `forest' of signatures is expected to be generated along each line of sight \cite{2015aska.confE.103G}.

\begin{figure}[H]
\centering
\includegraphics[width=13 cm]{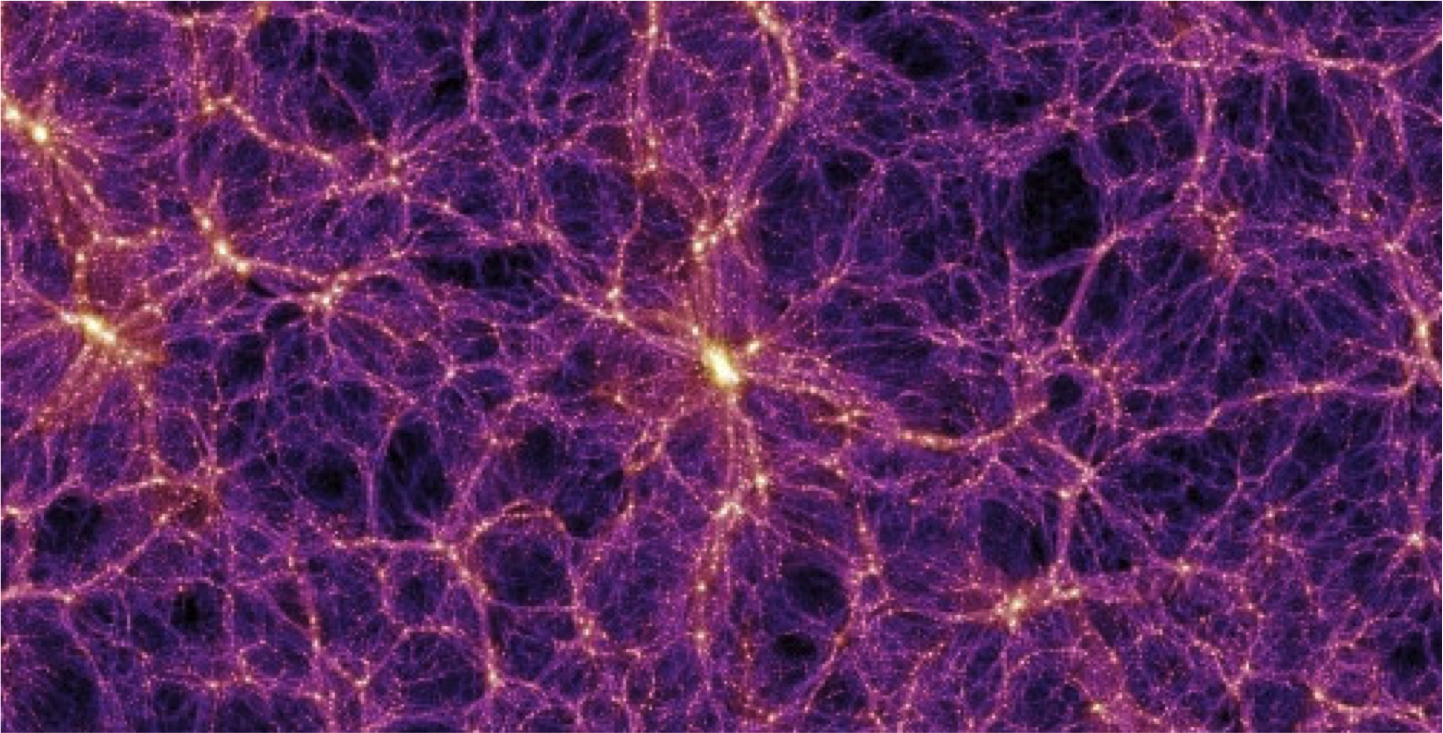}
\caption{The structure of the intergalactic medium can be illustrated by the Millenium simulation \cite{2005Natur.435..629S}. This ``cosmic web'' consists of filaments of galaxies and empty desolate voids which occupy the vast majority of the volume of the Universe. Measuring the elusive cosmic magnetic fields in this diffuse plasma is fundamentally dependent on the computationally-expensive technique of Faraday Tomography, which can have similar computational overheads to de-dispersion for observations of transients.
}
\label{cosmicweb}
\end{figure}

One of the key questions for magnetic field studies is to detect magnetism on the largest cosmic scales. At such scales, the ``cosmic web'' (shown in Fig.~\ref{cosmicweb}) consists of filaments of galaxies and empty desolate voids which occupy the vast majority of the volume of the Universe. While various predictions have been made as to the strength of these intergalactic magnetic fields \cite{2010ApJ...723..476A,2011ApJ...738..134A,2013ApJ...767..150A,2014ApJ...790..123A,2014PASJ...66...65A,2015HiA....16..407A,2015A&A...580A.119V,2016ApJ...824..105A,2016arXiv160207526V,2018arXiv180511113V}, the magnetic fields in this medium are so weak that they remain undetected observationally \cite{2017ApJ...841...67F}. One of the best techniques for measuring cosmic magnetic fields is Faraday Tomography, which can extract the Faraday rotation measure (RM) properties of extragalactic sources \cite{2014ApJS..212...15F} and thereby provides a unique probe of fundamental physics. For the purpose of measuring magnetism in the cosmic web, we require large statistical samples with highly precise Faraday rotation and polarized fraction measurements. Samples with $\sim40,000$ sources are currently available \cite{2009ApJ...702.1230T}, and the SKA itself is anticipated to generate up to 15~million measurements \cite{2015aska.confE.103G}. One way to explore an intermediate number of sources is by using currently available SKA pathfinder and precursor telescopes, of which one example is the LOw Frequency ARray (LOFAR; \cite{2013A&A...556A...2V}).

LOFAR is a huge data producer, with a raw data-rate of around 10~Tbit/s. Beam-forming at the station level is able to reduce the long-range data transport rate to around $\sim150$~Gbit/s, although this still requires dedicated fibre networks to transport the data. As an indication of scale, this is comparable to the bandwidth requirements of the whole of Netflix -- in 2015 the peak Netflix data pipeline rate was $192$~Gbit/s during peak hours. Following data transport, LOFAR data are processed at the University of Groningen by an IBM Blue Gene/P supercomputer offering about 30~Tflop/s of processing power \cite{2013A&A...556A...2V}. A key area at which the LOFAR telescope excels is in low-frequency radio surveys, and one survey that has been conducted is the Multifrequency Snapshot Sky Survey (MSSS; \cite{b2}). The polarized counterpart of this survey is the MSSS All-Sky Polarization Survey (MAPS), which will enable Faraday Tomography of the entire LOFAR sky. This survey is currently under development by the MSSS and MAPS teams (Heald et al., this volume), in addition to Southern sky counterparts such as POGS \cite{2018arXiv180909327R}. Surveys such as MAPS can be useful test-beds for science pipelines with the SKA, and for estimating the imaging performance of a spectral imaging pipeline -- which may potentially constitute one core component of the data processing requirements \cite[e.g.][]{2017MNRAS.465.3680M}. We have therefore been working alongside the MSSS/MAPS teams in developing a parallelised SKA spectral imaging pipeline, which can be tested upon LOFAR data.

MSSS/MAPS constitutes 50~terabytes of archived visibility data in measurement set format. The intention is to image the survey at 45~arcsec angular resolution. The survey has the goal of providing a direct detection of magnetic fields in the ``cosmic web'' of galaxy filaments and voids \cite[e.g.][]{2018arXiv180507995B}, of producing an ``RM grid'' at 150~MHz \cite[e.g.][]{2018A&A...613A..58V}, and also of furthering our understanding of extragalactic radio sources \cite[e.g.][]{2016ApJ...829....5L,2017MNRAS.469.4034O} -- with progress having recently been made in charting out the evolution of cosmic magnetic fields in galaxies, with intervening galaxy experiments now spanning a redshift range from $z=0.005$ \cite{2018MNRAS.476.2432G} to $z=2$ \cite{2017ApJ...841...67F}. MAPS will be able to provide up to $\approx200,000$~sources for such studies, with a conservative estimate of positive RM detections towards 100--200 sources due to strong Faraday depolarization effects at long wavelengths. The resulting catalogue of RMs, polarized fractions, and observational limits can be used to constrain the intergalactic magnetic field. This is only possible due to the high RM precision available with LOFAR, which enables Faraday Tomography with a rotation measure spread function (RMSF) -- the point spread function in Faraday space -- of FWHM $\sim1.3$~rad~m$^{-2}$ \cite{2013A&A...558A..72I,2013A&A...552A..58S,2014A&A...568A.101J,2014A&A...568A..74M,2015A&A...574A..73M,2015A&A...574A.114V,2016MNRAS.462.1910H,2017A&A...601A..25C,2017A&A...597A..98V}. As a bonus outcome of our science pipeline prototyping work, in addition to testing the deployment and operation of science for SKA, this project will also help to enable tight constraints on the magnetised large-scale structure of the Universe. The SKA therefore provides unique opportunities to contribute to the technical development of a scientific instrument which has considerable economic potential across numerous sectors, while simultaneously contributing to and enabling modern scientific discoveries by processing data from pre-existing pathfinder and precursor telescopes.

As a serialised pipeline, it had previously been estimated that MAPS would require up to 2~years of CPU time in order to just produce Stokes $Q$ \& $U$ images. However, due to the parallelisable and scalable computing infrastructure now available, the survey can now be completed on a feasible timescale. The scalability is primarily related to Dask itself, which has been tested on clusters containing thousands of cores. This same infrastructure could also be applied to other magnetism surveys, including for example, data from the Australia Square Kilometre Array Pathfinder (ASKAP) \cite{2007PASA...24..174J} and the Karl G. Jansky Very Large Array (VLA) \cite{2016AAS...22732409L}.

\section{An SKA-like Science Pipeline}
Faraday Tomography studies of magnetic fields require every single channel across the observing bandwidth to be imaged and CLEANed individually. In the case of SKA, this will require $65,000$ separate images in all Stokes parameters. This process lends itself to parallelism in a trivial manner, as it is possible to image each channel simultaneously on multiple cores distributed across a cluster in a high performance computing (HPC) environment.

One plausible method for parallelisation of this process is the Python-based Dask\footnote{\url{https://dask.pydata.org}}. Dask is open-source and freely available, with schedulers that scale to thousand-node clusters. Dask can, in some circumstances, bypass the Global Interpreter Lock (GIL) when using pure Python objects and numeric code such as NumPy or Scikit-Learn. In particular, the Dask distributed library enables a Python function to be submitted across a cluster for task-based parallelism, generating a number of futures that can be monitored, controlled, and computed as needed. The flexibility of Dask enables parallelism to be implemented in just a few lines of code. For example, the following code snippet sets up a Dask client, submits jobs across the cluster, and then monitors progress of the running futures. The $QU$ imaging function could, in principle, call any imaging code of interest to the user.

\begin{lstlisting}
import dask
from dask.distributed import Client
from distributed.diagnostics import progress
client = Client('localhost:8786')
futures = [client.submit(qu_imaging, data[chan]) for chan in range(65000)]

progress(futures) # show progress bar

def qu_imaging(data):
	# Run imaging code/software of choice.
\end{lstlisting}
Dask therefore represents one particular scalable method that we would like to further investigate for SKA-like science pipelines. The currently implemented science pipeline is purely Pythonic, and we have written our own interface using `Dask distributed' which submits and monitors the running imaging processes. The completed code will be open-source and available via GitHub, so that it is possible to download the code for application to one's favourite telescope of choice.

The science pipeline implements a suite of algorithms that are particularly specialised for the SKA. These include calibration of ionospheric Faraday rotation -- which has been previously studied for LOFAR \citep[e.g.][]{2016RaSc...51..927M} -- and can be corrected by using a total electron content model \cite{RMextract-ascl}. The pipeline also enables a complex CLEAN deconvolution for polarimetric data \cite{2016MNRAS.462.3483P}, the generation of Faraday Moment images to enable highly-complete source-finding \cite{2018MNRAS.474.3280F}, and finally the application of Faraday Tomography \cite{2005A&A...441.1217B}, together with a complex cross-correlation based RM-CLEAN \cite{2009IAUS..259..591H}. GPUs have already been widely deployed in the search for transients \cite{2012ASPC..461...33A}, and as part of our pipeline development work we have now also begun considering the capacity for deploying Faraday Tomography algorithms on GPUs in a manner that is scalable for SKA.

An initial parallelised version of this SKA-like science pipeline is now being tested. Preliminary results indicate that the pipeline significantly reduced the time required to process a single pointing of sky. For a test LOFAR dataset, a serialised code took $11$~hours to complete, whereas a parallelised Dask pipeline required just $8$~minutes. In addition, source parameterisation measurements (sky coordinates and flux density) of a pulsar detected using these test datasets have been found to be consistent with those from other radio astronomy packages, thereby confirming the technical accuracy and scientific veracity of the science pipeline.

\section{Conclusions}
Our development work will test the scalability and the implementation of SKA science pipelines. We are aiming to test a fully working and deployable spectral imaging and Faraday Tomography pipeline by the end of 2018.

Our prototyping work also has multiple additional outcomes. It allows us to optimise and validate the scalability of the Faraday Tomography algorithm in the SKA-era. It allows us to provide a realistic SKA-like data pipeline, that can be optimised for, deployed, and tested on HPC infrastructure. It also provides bonus outcomes: in terms of generating all-sky datasets that can be provided for science publications, and the potential discovery of the magnetised large-scale structure of the Universe.

\vspace{6pt} 



\authorcontributions{Conceptualisation, J.F., B.M., and W.A.; Methodology, J.F., B.M., F.D., S.S., and W.A.; Software, J.F., B.M., F.D., S.S., and W.A.; Validation, J.F., B.M., and F.D.; Formal Analysis, J.F.; Writing—Original Draft Preparation, J.F.; Writing—Review \& Editing, J.F.; Visualisation, J.F.; Supervision, B.M. and W.A.; Project Administration, W.A.; Funding Acquisition, W.A. Authorship is limited to those who have contributed substantially to the work reported.}

\funding{This work is supported by STFC Grants: SKA Preconstruction funding supplement (ST/N003713/1), SKA Preconstruction update (ST/P005446/1) and SKA Preconstruction 2017-18 (ST/R000557/1).}

\acknowledgments{We are grateful to the wider Scientific Computing group at OeRC, including Karel Ad\'{a}mek, Anna Brown, and Jan Novotny, for support and insightful comments that helped to improve this document. We are also grateful to the MSSS/MAPS teams, led by George Heald and Jess Broderick, for allowing us to test these pipelines with LOFAR data. LOFAR, designed and constructed by ASTRON, has facilities in several countries, that are owned by various parties (each with their own funding sources), and that are collectively operated by the International LOFAR Telescope (ILT) foundation under a joint scientific policy.}

\conflictsofinterest{The authors declare no conflict of interest.} 

\abbreviations{The following abbreviations are used in this manuscript:\\

\noindent 
\begin{tabular}{@{}ll}
ASKAP & Australia SKA Pathfinder\\
CPU & Central Processing Unit\\
GIL & Global Interpreter Lock\\
GPU & Graphics Processing Unit\\
HPC & High Performance Computing\\
LOFAR & LOw Frequency ARray \\
MAPS & MSSS All-Sky Polarization Survey\\
MSSS & Multifrequency Snapshot Sky Survey\\
POGS & POlarised GLEAM Survey\\
RFI & Radio-Frequency Interference\\
RM & Rotation Measure\\
RMSF & Rotation Measure Spread Function\\
SKA & Square Kilometre Array\\
STFC & Science and Technology Facilities Council\\
VLA & Karl G. Jansky Very Large Array
\end{tabular}}


\reftitle{References}





\end{document}